\documentclass[twocolumn]{revtex4-1}
\usepackage{bm}
\usepackage{amsmath}
\usepackage{amssymb}
\usepackage{graphicx}

\def\0{{\bf 0}}
\def \E {{\cal E}}
\def \LL {{\cal L}}

\def\b#1{{\mathbb #1}}
\def\nn{\nonumber \\}

\newcommand{\bY}{{\bf Y}}
\newcommand{\bx}{{\bf x}}

\newcommand{\bxp}{{\bf x}^{\scriptscriptstyle \perp}}
\newcommand{\hbxp}{\hat {\bf x}^{\scriptscriptstyle \perp}}
\newcommand{\bX}{{\bf X}}
\newcommand{\bu}{{\bf u}}
\newcommand{\bup}{{\bf u}^{\scriptscriptstyle \perp}}
\newcommand{\hbup}{\hat {\bf u}^{\scriptscriptstyle \perp}}

\newcommand{\bw}{{\bm w}}

\newcommand{\bv}{{\bf v}}

\newcommand{\rx}{{\rm x}}
\newcommand{\Ba}{{\bm \alpha}}
\newcommand{\Bap}{{\bm \alpha}^{\scriptscriptstyle \perp}}
\newcommand{\bb}{{\bm \beta}}

\newcommand{\Be}{{\bm \epsilon}}
\newcommand{\Bep}{{\bm \epsilon}^{\scriptscriptstyle \perp}}
\newcommand{\bE}{{\bf E}}
\newcommand{\bEp}{{\bf E}^{\scriptscriptstyle \perp}}
\newcommand{\hbEp}{\hat{\bf E}^{\scriptscriptstyle \perp}}

\newcommand{\bep}{{\bf e}^{\scriptscriptstyle \perp}}
\newcommand{\bbp}{{\bf b}^{\scriptscriptstyle \perp}}
\newcommand{\bB}{{\bf B}}
\newcommand{\bBp}{{\bf B}^{\scriptscriptstyle \perp}}
\newcommand{\hbBp}{\hat{\bf B}^{\scriptscriptstyle \perp}}

\newcommand{\bK}{{\bf K}}
\newcommand{\bKp}{{\bf K}^{\scriptscriptstyle \perp}}
\newcommand{\bA}{{\bf A}}
\newcommand{\bAp}{{\bf A}^{\scriptscriptstyle \perp}}

\newcommand{\bjp}{{\bf j}^{\scriptscriptstyle \perp}}

\newcommand{\Bp}{{\bf p}}
\newcommand{\bP}{{\bf P}}
\newcommand{\bPi}{{\bf \Pi}}

\newcommand{\bi}{\mathbf{i}}
\newcommand{\bj}{\mathbf{j}}
\newcommand{\bk}{\mathbf{k}}

\newcommand{\be}{\begin{equation}}
\newcommand{\ee}{\end{equation}}
\newcommand{\bea}{\begin{eqnarray}}
\newcommand{\eea}{\end{eqnarray}}
\newcommand{\ba}{\begin{array}}
\newcommand{\ea}{\end{array}}
\begin{document}
\title{On a `time' reparametrization in relativistic electrodynamics with travelling waves}

\author{  Gaetano Fiore$^{1,2}$
   \\    
$^{1}$ Dip. di Matematica e Applicazioni, Universit\`a di Napoli ``Federico II'',\\
Complesso Universitario  M. S. Angelo, Via Cintia, 80126 Napoli, Italy\\         
$^{2}$         INFN, Sez. di Napoli, Complesso  MSA,  Via Cintia, 80126 Napoli, Italy}

\begin{abstract}
We briefly report on our method \cite{Fio17} of
 simplifying the equations of motion of  charged particles in an electromagnetic (EM) field that is the sum of a plane travelling wave  and a static part; it is based
on changes of the  dependent variables and the independent one (light-like coordinate $\xi$ instead of time $t$). We sketch its application to a few  cases of extreme laser-induced accelerations, both in vacuum and in plane problems at the vacuum-plasma interface, where we are able to reduce the system of the (Lorentz-Maxwell and continuity) partial differential equations into a family of decoupled systems of Hamilton equations in 1 dimension. 
Since Fourier analysis plays no role, the method can be applied to all kind of travelling waves, ranging from almost monochromatic to socalled “impulses".
\end{abstract}

\maketitle
\section{Introduction and set-up}
\label{intro}

The equation of motion  of a particle with charge $q$ in  external electric and magnetic  fields 
$\bE(\rx),\bB(\rx) $  [$\rx\!\equiv\!(ct,\bx)$] 
 in its general form is non-autonomous and highly nonlinear:
\bea
\ba{l}
\dot\Bp(t)=q\bE[ct,\bx(t)] + q\bb(t) \wedge \bB[ct,\bx(t)] ,\\[6pt]  
\displaystyle
\dot \bx(t) =\frac{c\Bp(t) }{\sqrt{m^2c^2\!+\!\Bp^2(t)}} ;
\ea
\label{EOM}
\eea
here $\bb\!\equiv\!\bv/c$, 
$\Bp\!\equiv\! m\bv/\sqrt{1\!-\!\bb^2}$ is its relativistic momentum.

\medskip
Usually, (\ref{EOM}) is  simplified assuming:

\begin{enumerate}

\item $\bE,\bB$ are constant or vary  ``slowly'' in space/time; \ or

\item $\bE,\bB$  are  ``small" (so that nonlinear effects  in $\bE,\bB$ are negligible);  \ or

\item $\bE,\bB$ are monochromatic waves, or slow modulations of;   \ or

\item the  motion of the particle keeps non-relativistic.

\end{enumerate}

The astonishing developments of  Laser technologies   (especially {\it Chirped Pulse
Amplification} \cite{StriMou85,MouTajBul06}) today allow the construction of compact
sources of extremely intense (up to $10^{23}\,$W/cm$^2$) coherent EM waves,
possibly concentrated in very short laser pulses ($\gtrsim$ fs).
Even more intense/short (or cheaper) laser pulses by new technologies
(thin film compression \cite{MouMirKhaSer14}, etc.) will be soon available.
In particular, these lasers can be used for making small  particle-accelerators based 
on Laser Wake Field Acceleration (LWFA) \cite{Tajima-Dawson1979} in plasmas. 
Extreme conditions are present also in several violent astrophysical processes
(see e.g. \cite{TajNakMou17} and references therein).
In either case the effects are so fast, huge, highly nonlinear, ultra-relativistic
that conditions 1-4  are not fulfilled. Alternative simplifying approaches are
therefore desirable.

\medskip
Here we summarize a new approach \cite{Fio17} that is especially fruitful if in the spacetime region $\Omega$ of interest
(i.e., where we wish to follow the charged particles' worldlines) $\bE,\bB$  can be decomposed into
a static part and a plane transverse travelling wave propagating in the $z$ direction:
\bea
\ba{lcl}
\bE(\rx)=&\underbrace{\Bep(ct\!-\!z)}_{pump = travelling\, wave}&+\underbrace{\bE_s(\bx)}_{static},\\[20pt] \bB(\rx)=&\underbrace{\bk\wedge\Bep(ct\!-\!z)}_{travelling\, wave} &+
\underbrace{\bB_s(\bx)}_{static},
\ea
\label{EBfields}
\eea
$\bx\!=\!x\bi\!+\!y\bj\!+\!z\bk$, $\Bep\!\!\perp\!\bk$. We decompose vectors as $\bu\!=\!\bup\!+\!u^z\bk$.
We assume {\it only}  that $\Bep(\xi)$ is piecewise continuous and
\bea
\ba{ll}
&\mbox{ {\bf a}) }\quad \Bep \mbox{ has a compact support }[0,l],\\[10pt] 
\mbox{ or} &\mbox{ {\bf a'})}\quad  \Bep \in L^1(\mathbb{R}),
 \ea   \label{aa'} \\
\Rightarrow\:\:\Bap(\xi)\!\equiv\!-\!\!\int^{\xi}_{ -\infty }\!\!\!\!\!\!\!dy\,\Bep\!(y) 
\!\to\! 0\quad \mbox{as }\xi\!\to\!-\infty;         \label{defBap}
\eea
$\Bap$ is the travelling-wave part of the transverse EM potential $\bAp$.
{\bf a}) $\Rightarrow$ $\Bap(\xi)\!=\!0$ if $\xi\!\le\! 0$,  $\Bap(\xi)\!=\!\Bap(l)$ if $\xi\!\ge\! l$. \
We can treat on the same footing all such $\Bep$, in particular:
\begin{enumerate}
\item A modulated monochromatic wave fulfilling (\ref{aa'}):
\be
\Bep\!(\xi)\!=\!\underbrace{\epsilon(\xi)}_{\mbox{modul.}}
\underbrace{[\bi a_1\cos (k\xi\!+\!\varphi)\!+\!\bj a_2\sin (k\xi)]}_{\mbox{carrier wave $\Be_o^{{\scriptscriptstyle \perp}}\!(\xi)$}}
 \label{modulate}
\ee
\be
\Rightarrow\:\:\: -\Bap(\xi)\!=\!   \frac {\epsilon(\xi)}{k^2} \,\Bep_o{}'\!(\xi)+O\left(\frac 1 {k^2}\right)
 \!\simeq\!   \frac {\epsilon(\xi)}{k^2} \,\Bep_o{}'\!(\xi);      \label{slowmodappr}
\ee
$\simeq$ holds if \ $|\epsilon'|\!\ll\! |k\epsilon|$ \ (slow modulation).
\label{modula1}

\item 
A superposition of waves of type 1.
\label{modula2}

\item An `impulse'  (few cycles, or even a fraction of).

\end{enumerate}

\begin{figure}
\centering
\includegraphics[width=8cm]{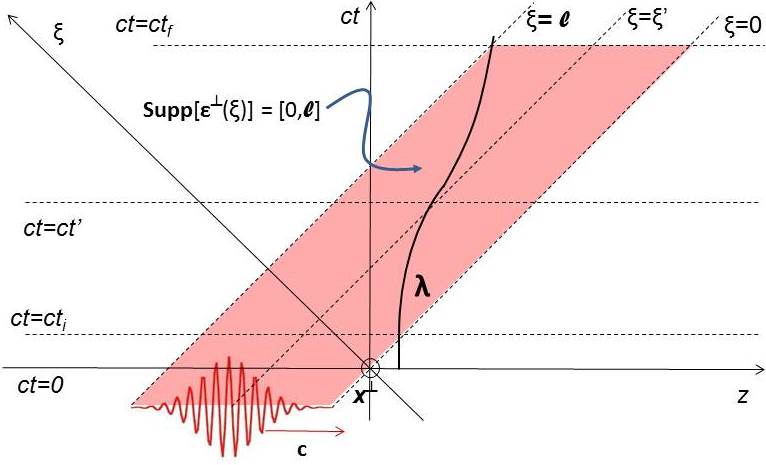}
\caption{Every worldline $\lambda$  and hyperplane $\xi\!=$C
intersect once.}
\label{Worldline}      
\end{figure}

\noindent
The idea  is: as no particle can 
reach the speed of light $c$, \ $\tilde \xi(t)\!=\!ct\!-\!z(t)$ is strictly growing,
and we can adopt $\xi\!=\!ct\!-\!z$ as a parameter on the worldline $\lambda$ 
(see fig. \ref{Worldline}) and in the action functional of the particle:  
\bea
S(\lambda) = -\!\int_\lambda\! mc^2  d\tau+ qA(\rx) \qquad\qquad\qquad\qquad\nn
= -\!\int\limits_{t_0}^{t_1}\!\!dt\,\underbrace{\frac{mc^2\!\!+ \!q  u^\mu A_\mu}{\gamma}}_{ L[\bx,\dot\bx,t]}
=-\!\int\limits_{\xi_0}^{\xi_1}\!\!\frac{d\xi}c\,\underbrace{\frac{mc^2\!\!+ \!q \hat u^\mu \hat A_\mu }{\hat s}}_{ \LL[\hat \bx,\hat \bx',\xi]};\nonumber
\eea 
 $(u^\mu)\!=\!(u^0\!,\bu)\!\equiv\!(\gamma,\gamma \bb)
$ is the 4-velocity, i.e. the dimensionless version of the 4-momentum. 
$A(\rx)\!=\!A_\mu(\rx)d\rx^\mu\!=\!A^0(\rx)cdt\!-\!\bA(\rx)\!\cdot\! d\bx$ \ is the EM potential 1-form, \ $\bE=-\partial_t\bA/c-\nabla A^0$,
 $\bB=\nabla\!\wedge\!\bA$ (we use
Gauss CGS units), and we denote $\hat \bx(\xi)=\bx(t)$,   $\hat f(\xi, \!\hat \bx)\!\equiv\! f(ct,\!\bx)$,   
$\hat f'\!\equiv\! d\hat f/d\xi$ for all functions $f(ct,\!\bx)$.
 Applying Hamilton's principle and the Legendre transform
we find simplified Lagrange, and Hamilton equations where  the argument of $\Bep$ 
is the independent variable $\xi$, rather than  the unknown $ct\!-\!z(t)$, and the new kinetic momenta are the dependent variables.

\section{General results for one particle}
\label{sec-1}

To parametrize $\lambda$ by $\xi$ we have to replace \
$d\tau/dt\!=\!1/\gamma\!=\!\sqrt{1\!-\! \dot\bx^2/c^2}$ \ ($\tau$ is the particle proper time) by
\be
\frac 1{\hat s}\equiv \frac {d (c\tau)}{d\xi}=
\sqrt{1\!+\!2\hat z'\!-\!\hat\bx^{\scriptscriptstyle \perp}{}'{}^2} \:\: >0. \label{defs}
\ee
From  \ $\Bp =m d\bx/d\tau$, $\gamma=dt/d\tau$ \ we find that the   {\it $s$-factor} $\hat s$ is
 the light-like component \ $\hat u^-=\hat \gamma\!- \hat u^z$ \ of the 4-velocity $u\!=\!(u^0\!,\bu)
\!\equiv\!(\gamma,\!\gamma \bb)\!=\!\left(\!\frac {p^0}{mc^2},\!\frac {{\bf p}}{mc}\!\right)$ 
(all these are dimensionless), 
and  $\hat\bu \!=\!\hat s \hat \bx'$. \
$\hat\gamma,\hat u^z,\hat\bb,\hat \bx'$ can be expressed as  
{\it rational  functions} of $\hat\bu^{{\scriptscriptstyle\perp}},\hat s$:
\bea
&&\hat\gamma\!=\!\frac {1\!+\!\hat\bu^{{\scriptscriptstyle\perp}}{}^2\!\!+\!\hat s^2}{2\hat s}, 
\quad \hat u^z\!=\!\hat\gamma\!-\!\hat s, 
 \quad \hat\bb\!=\! \frac{\hat\bu}{\hat\gamma},\quad \label{u_es_e} \\
 &&\hbxp{}'
=\displaystyle\frac {\hbup}{\hat s}, \qquad \quad  \hat z'
=\displaystyle\frac {1\!+\!\hbup{}^2}{2\hat s^2}\!-\!\frac 12 \label{eqx}
\eea
By Hamilton's principle, any extremum $\lambda$ of $S$  is the worldline of a possible motion of the particle with initial position $\bx_0$ at time $t_0$ and final position $\bx_1$ at time $t_1$. Hence it  fulfills Euler-Lagrange equations in both forms \
$\frac d {dt}\frac{\partial L}{\partial \dot \bx}=\frac{\partial L}{\partial \bx}$ and
$\frac d {d\xi}\frac{\partial \LL}{\partial \hat \bx'}=\frac{\partial \LL}{\partial \hat \bx}$,  \ equivalent to (\ref{EOM}).
The Legendre transform  yields the Hamiltonians $H\!\equiv\! \dot\bx\!\cdot\!\frac{\partial L}{\partial \dot \bx}\!-\!L\!=\!\gamma mc^2\!+\!q  A^0$ and  \ 
$\hat H\!\equiv\!\hat \bx'\!\cdot\!\frac{\partial \LL}{\partial \hat \bx'}\!-\!\LL\!=\!\hat\gamma mc^2\!+\!q\hat A^0$. \
$\hat H$ is a {\it rational} function of $\hat \bx,\hat \bPi\!\equiv\!\frac{\partial \LL}{\partial \hat \bx'}$, or, equivalently, of $\hat s,\hat\bu^{{\scriptscriptstyle\perp}}$:
\bea
&&\hat H(\hat \bx,\!\hat\bPi;\!\xi)=mc^2\frac {1\!+\! \hat s^2\!\!+\! \hat \bu^{{\scriptscriptstyle\perp}2}\!}{2\hat s}\! +\! q\hat A^0\!(\xi,\hat\bx),\qquad
\label{Ham} \\ &&\qquad\mbox{where}\:\: 
 \left\{ 
\ba{l} \displaystyle mc^2\hat\bu^{{\scriptscriptstyle\perp}}\!\!=\!\hat\bPi^{{\scriptscriptstyle\perp}}\!
\!-\!q\hat \bA^{{\scriptscriptstyle\perp}}(\xi,\hat\bx),\\[6pt]
\displaystyle mc^2\hat s\!=\!-\hat \Pi^z\!\!-\!q\hat A^-(\xi,\hat\bx),        
\ea \right.     
\eea
 while \ $H(\bx,\bP,t)\!=\!\sqrt{\!m^2c^4\!+\!(c\bP\!-\!q\bA)^2}+\!qA^0$ 
\ ($\bP\!\equiv\!\frac{\partial L}{\partial \dot \bx}\!=\!\Bp\!+\!\frac qc\bA$) is not.
Eq. (\ref{EOM})  are also equivalent to the Hamilton equations \
$\hat \bx'\!=\!\frac{\partial \hat H}{ \partial \hat\bPi},\: \hat\bPi' \!=\!-\frac{\partial \hat H}{ \partial \hat \bx}$. 
All the new equations (in particular these ones) can be also obtained more  directly from the old ones
by putting a caret on all dynamical variables and replacing \ $d/dt$ by $(c\hat s/\hat \gamma)d/d\xi$.
Along the solutions  $\hat H$ gives the particle energy  as a function of $\xi$.
Under the EM field  (\ref{EBfields}) eqs  (\ref{EOM}) amount to (\ref{eqx}) and
\bea
\ba{l}\displaystyle\hbup{}'\!=\frac q{mc^2}\!\left[(1\!+\!\hat z')\hbEp_s\!+\!(\hat\bx'\!\wedge\!\hat\bB_s)^{\scriptscriptstyle \perp}\!+\!\Bep(\xi) \right]\!,\\[10pt]
\displaystyle\hat s'=\frac {-q}{mc^2}\left[\hat E^z_s-\hbxp{}'\!\cdot\!\hbEp_s\!+(\hbxp{}'\!\wedge\!\hbBp_s)^z\right]\!,    
\ea\qquad \label{equps} 
\eea
while the  {\it energy gain}  (normalized to $mc^2$) is 
\be
\E\equiv \frac{\hat H(\xi_1)\!-\!\hat H(\xi_0)}{mc^2}= \int^{\xi_1}_{\xi_0}\!\!\! d\xi\,q\Bep\!\cdot\!\frac{\hbup}{\hat s }
\label{EnergyGain}
\ee 
in the interval $[\xi_0,\xi_1]$.
Once solved (\ref{eqx}-\ref{equps}), analytically or numerically, to obtain the solution as a function of $t$
we just need to invert $\hat t (\xi)\!=\!\xi\!+\!\hat z(\xi)$ and set $\bx(t)=\hat\bx[\xi(t)]$.
If $\bE_s,\bB_s\!=$const then
eq. (\ref{equps}) are solved by
\bea
\ba{l}
\hbup=\displaystyle\frac q{mc^2}\left[\bK\!-\Ba(\xi)\!+\!(\xi\!+\!\hat z)\bE_s\!+\!
\hat\bx\!\wedge\!\bB_s \right]^{\scriptscriptstyle \perp}\!\!,\\[10pt]
\hat s=\displaystyle\frac {-q}{mc^2}\left[K^z\!+\xi E^z_s-\hbxp \!\cdot\!\bEp_s\!+(\hbxp\!\wedge\!\bB_s)^z\right]
\ea    \label{constEsBs}
\eea
($K^j $ are integration constants) whereby (\ref{eqx}) become three  1$^{st}$
order ordinary differential equations (ODE)  {\it rational} in the unknown $\hat \bx(\xi)$.

\smallskip
Contrary to  (\ref{eqx}-\ref{equps}),  (\ref{EOM}) is a transcendental system, and
the unknown $z(t)$ appears
in the argument of the rapidly varying functions $\Bep,\Bap$ in (\ref{EOM})$_1$, which now reads:
\bea
\ba{l}
\frac 1q \dot\Bp(t)\!=\! \bE_s\!\!+\!\bb\!\wedge\!\bB_s\!+\! \Bep[ct\!-\!z(t)]\left(\!\cdot\bb\bk\!+\!1\!-\!\beta^z\right)\!.
\ea
\nonumber 
\eea
Also determining $\E(t)$ is more complicated.

\subsection{\small Dynamics under \  $A^\mu\!=\!A^\mu(t,\!z)$}
\label{A=A(t,z)}

This applies in particular to (\ref{EBfields}) if $\bE_s= E_s^z(z)\bk$, $\bB_s=\bB_s^{\scriptscriptstyle \perp}(z)$, choosing
e.g. \ $A^0=-\int^z\!d\zeta E_s^z(\zeta)$, \ $\bAp\!=\Bap\!-\!\bk\wedge\!\int^z\!d\zeta \bBp(\zeta)$, $A^z\!\equiv\! 0$.
As $\partial \hat H/\partial \hat \bx^{\scriptscriptstyle \perp}\!=\!0$, we find
$\hat\bPi^{\scriptscriptstyle \perp}\!=\!q\bKp\!=\!\mbox{const}$, i.e.
the known result $\frac {mc^2}q\hat\bu^{\scriptscriptstyle \perp}\!=\!\bKp
\!-\!\hat \bA^{{\!\scriptscriptstyle\perp}}\!(\xi,\!\hat z)$.
  Setting $v\!:=\!\hbup{}^2$  and replacing  in  (\ref{eqx})$_2$, (\ref{equps})$_2$ we obtain 
\bea
\hat z'=\displaystyle\frac {1\!+\!\hat v}{2\hat s^2}\!-\!\frac 12, \quad
\hat s'=\frac{-q}{mc^2}E_s^z(\hat z)\!-\!\frac 1{2\hat s}
\frac{\partial \hat v}{\partial \hat z}.     \label{reduced}
\eea

Once solved the system (\ref{reduced})  in the unknowns $\hat z(\xi), \hat s(\xi)$,
the other unknowns are  obtained from 
\bea
&& \hat\bx(\xi)=\bx_0+\hat \bY\!(\xi), \qquad  \hat \bY\!(\xi)\!\equiv\!\!\displaystyle\int^\xi_{\xi_0}\!\!\! dy \,\frac{\hat\bu(y)}{\hat s(y)}.\qquad \label{hatsol}
\eea
\smallskip
If in addition $\bB_s\!\equiv\!0$, then 
 $\bA_s\!\equiv\!0$  (in the Coulomb gauge),
\ $\hat\bu^{\scriptscriptstyle \perp}\!(\xi)\!=\!\frac q{mc^2}\left[\bK^{\scriptscriptstyle \perp}\!-\Bap(\xi)\right] $ and $ \hat v\!=\!\hat\bu^{{\scriptscriptstyle\perp}2}$
 are already known. The system   (\ref{reduced}) to  be solved simplifies to
\bea
&& \hat z'=\frac {1\!+\! \hat v}{2\hat s^2}\!-\!\frac 12, \qquad  \hat s'=\frac{-q}{mc^2}E_s^z(\hat z).
 \label{heq1r} 
\eea
Some remarkables properties of the solutions are  \cite{Fio17}:  

\begin{enumerate}

\item Where $\Bep(\xi)\!=\! 0$ then $ \hat v(\xi)\!=\!v_c\!=$const,
$\hat H$ is conserved,  (\ref{heq1r}) is solved by quadrature.

\item 
\label{notransv}
The final transverse momentum is \ $mc\bup(\xi_f)$.
 If  $\epsilon$ of (\ref{modulate}) varies slowly and $\bup\!(0)\!=\!\0$, then $\bup(\xi_f)\!\simeq\!0$.

\item 
$\hat s(\xi)$  is insensitive to fast  oscillations of  $\Bep$, contrary to $\bu,\gamma,\bb$, 
which can be reobtained via (\ref{u_es_e}).

\end{enumerate}

\section{Some exact solutions for $\bB_s,\bE_s\!=$const}

Let  $\bbp\!+\!b\bk\! \equiv\! q\bB_s/mc^2$, 
$\bep\!\equiv\! q\bEp_s/mc^2$ (constants), $\bw(\xi)\! \equiv\! q\!\left[\bK\!\!-\!\Bap(\xi)\!+\xi\bE_s\right]\!/\!mc^2$
(all dimensionless); (\ref{constEsBs}) take the more explicit form
\bea
\ba{l}
\hat u^x= (e^x\!-\!b^y)\hat z\!+\!b \hat y\!+\! w^x(\xi), \\[6pt]
\hat u^y= (e^y\!+\!b^x)\hat z\!-\!b \hat x\!+\! w^y(\xi), \\[6pt]
\hat s=(e^x\!-\!b^y)\hat x \!+\!(e^y\!+\!b^x)\hat y\!-\!w^z(\xi),
\ea    \label{constEsBs'}
\eea
For any $E^z_s,B^z_s,\bEp_s$, if  $\bBp_s\!=\!\bk\wedge\bEp_s$,  \ setting $\kappa\!\equiv\!\frac{qE_s^z}{mc^2}$
we find the following exact solutions (part of them are new):
 \bea
\ba{l}
\displaystyle ( \hat x +i \hat y)(\xi)=(1\!-\!\kappa\xi)^{ib/\kappa}\int^\xi_0\!\!\!\!  d\zeta \,
 \frac{(w^x+iw^y) (\zeta)} {(1\!-\!\kappa\zeta)^{1+ib/\kappa}}, \\[12pt]
\displaystyle   \hat z(\xi)\!=\!\!\int\limits^\xi_0\!\!  \frac {d\zeta }{2}\!
\left[\frac {1}{(1\!-\!\kappa\zeta)^2}\!+\!\hbxp{}'{}^2\!(\zeta)\!-\!1\right]\!, \:\:
\hat s(\xi)\!=\!1\!-\!\kappa\xi, \\[12pt]
 \hbup(\xi)\!=\!(1\!-\!\kappa\xi)\,\hbxp{}'(\xi),  \quad
  \hat \gamma(\xi)\!=1\!-\!\kappa\xi\!+\!\hat u^z(\xi)                      \\[14pt]
\displaystyle            \hat u^z(\xi)\!=\!\frac {1}{2(1\!-\!\kappa\xi)}+
(1\!-\!\kappa\xi)\,\frac {\hbxp{}'{}^2(\xi)\!-\!1}{2};
\ea   \label{SolEqBzEz}
\eea
 here we have adopted the initial conditions $\bx(0)\!=\!\0\!=\!\bu(0)$.
We next analyze a few special cases.

\subsection{Case $\bE_s\!=\!\bB_s\!=\!\0$ (zero static fields)}

Then (\ref{SolEqBzEz}) becomes \cite{LanLif62,Fio14JPA}:
\bea
\ba{l}
\displaystyle\hat s\!\equiv\! 1, \quad  \hat\bu^{\scriptscriptstyle \perp}\!\!=\!
\frac {-q\Bap}{mc^2}, \quad \hat u^z\!=\!\frac {\hat  \bup{}^2}2, \quad\hat\gamma\!=\!1\!+\!\hat u^z\\[16pt] 
\displaystyle \hat z(\xi)\!=\! \int ^\xi_{\xi_0}\!\!\!\!dy\,\frac{\hat\bu^{{\scriptscriptstyle\perp}2}(y)}2,\quad \hbxp\!(\xi)\!=\! \int ^\xi_{\xi_0}\!\!\!\!dy\,\hbup\!(y).      \ea                      \label{U=0s=0}
\eea
The solutions (\ref{U=0s=0}) induced by two $x$-polarized pulses 
and the corresponding $e^-$ trajectories  in the $zx$ plane  are shown in fig. \ref{Ez=0}. Note that:
\begin{figure*}
\centering
\includegraphics[width=8cm]{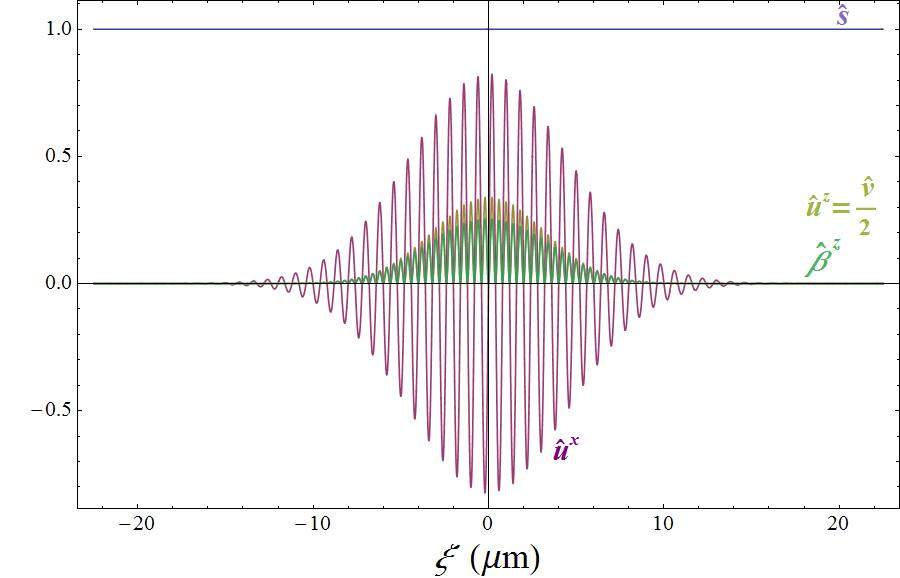} \hfill  \includegraphics[width=8cm]{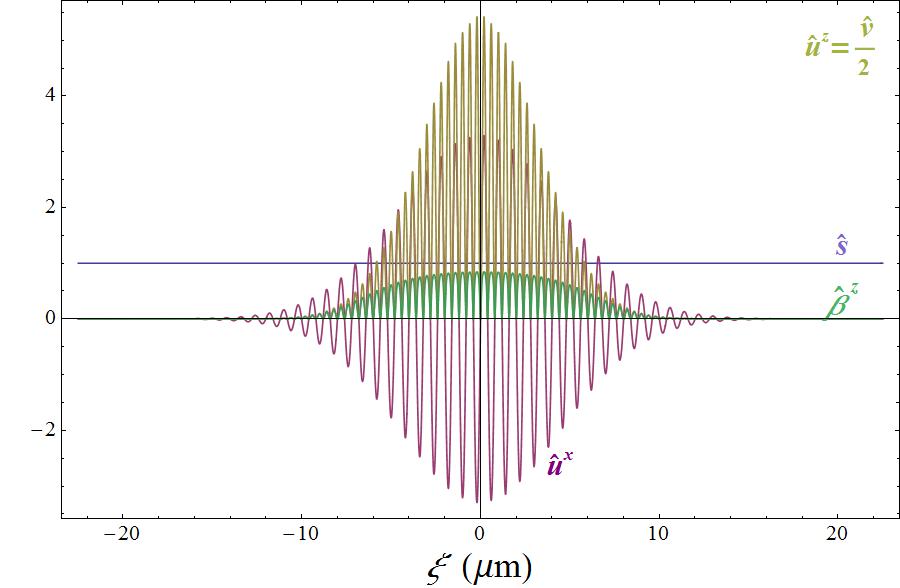}\\
\includegraphics[width=8cm]{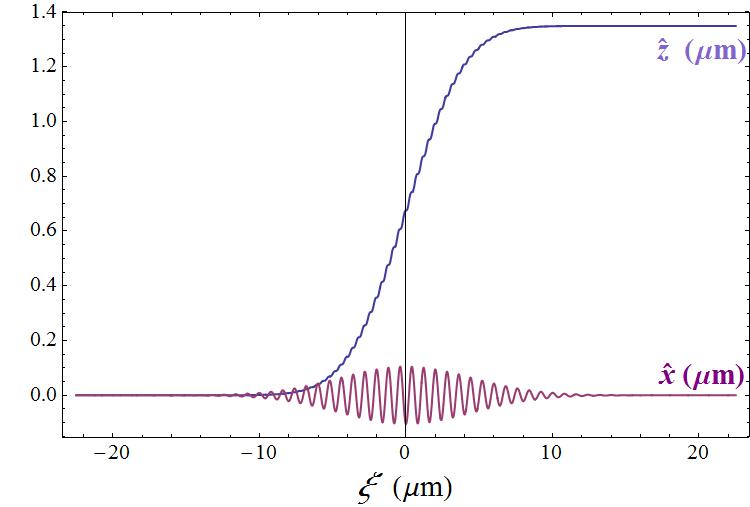}\hfill \includegraphics[width=8cm]{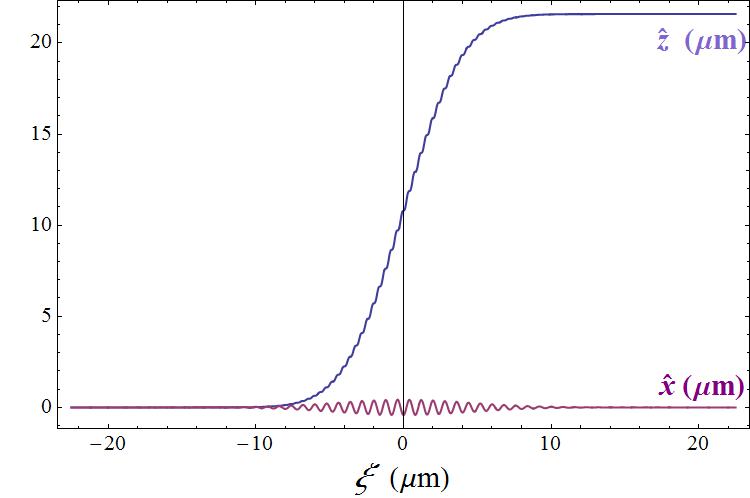} \\
\vskip.3cm
\includegraphics[width=5.8cm]{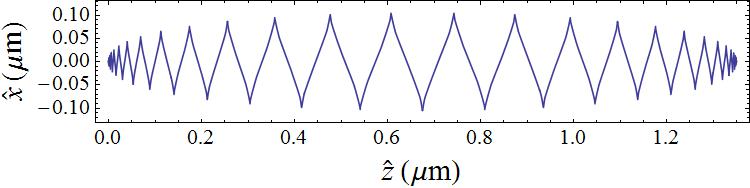} \hfill 
\includegraphics[width=10.2cm]{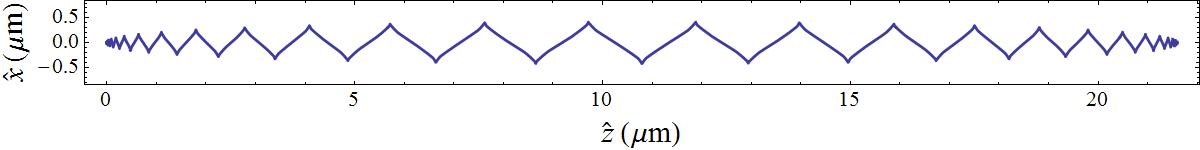}
\caption{Solutions (\ref{U=0s=0}) and $e^-$ trajectories  in the $zx$ plane induced 
by two $x$-polarized pulses with 
carrier wavelength $\lambda\!=\!.8\mu$m, gaussian modulation $\epsilon(\xi)=a\exp[-\xi^2/2\sigma]$,
$ \sigma \!=\! 20\mu$m$^2$, \ $e a\lambda/mc^2  \!=\! 4, 15$ (left, right). 
}
\label{Ez=0}      
\end{figure*}

$\bullet$ \ The maxima of $\gamma$, $\alpha^{\scriptscriptstyle \perp}$ 
coincide (and approximately also of $\epsilon(\xi)$, if $\epsilon(\xi)$ is slowly varying).

\smallskip
$\bullet$ \ Since $u^z\!\ge\!0$,  the $z$-drift is positive-definite.
Rescaling $\Bep\mapsto a\Bep$, $\hbxp,\hat \bup$ scale like $a$, whereas  $\hat z,\hat u^z$  scale like $a^2$
(hence the trajectory goes to a straight line in the limit $a\!\to\!\infty$). 
This is due to  magnetic force $q\bb \wedge \bB$.

\smallskip
$\bullet$  \ {\bf Corollary} \
The final $u$ and energy gain  read
\be
\bup_f\!=\!\hbup(\infty), \quad u^z_f=\E_f= \frac 12\bup_f{}^2=\gamma_f\!-\! 1;
 \label{Lawson-1} 
\ee
Both are very small if the pulse modulation $\epsilon$ is slow [extremely small if 
\ $\epsilon\!\in\!{\cal S}(\mathbb{R})$ \ or \ $\epsilon\!\in\!C^\infty_c(\mathbb{R})$].

Recall the {\it Lawson-Woodward}  
Theorem \cite{Law84,Pal88,
Pal95,EsaSprKra95} 
 (an outgrowth of the
original Woodward-Lawson Theorem  \cite{Woo47,WooLaw48}): 
 in spite of large energy variations during the interaction, 
the final energy gain $\E_f$ of a charged particle ${\cal P}$ interacting
with an EM field is zero if: 

i)  the interaction occurs in $\b{R}^3$ vacuum (no boundaries); 

 ii)  $\bE_s=\bB_s=\0$  and $\Bep$ is slowly modulated;  

iii)  $v^z\simeq c$ along the whole acceleration path; 

 iv) nonlinear  (in $\Bep$) effects $q\bb\!\wedge\! \bB$ are negligible; 

v) the power radiated by ${\cal P}$ is negligible. 

\noindent 
Our Corollary, as Ref. \cite{TrohaEtAl99}, states the same result if we relax iii), iv),
but the EM field is a {\it plane} travelling wave. 

\smallskip
To obtain a non-zero $\E_f$ one has to violate some other  conditions 
of the theorem, as e.g. we see in next cases.

\subsection{Case \ $\bE_s=0$, \ $\bB_s=B^z_s\bk $}

Then (\ref{SolEqBzEz}) becomes \ $\hat s\equiv 1$ and
\bea
\ba{l}
\displaystyle  ( \hat x \!+\! i \hat y)(\xi)\!=\!\!\int^\xi_0\!\!\!\!  d\zeta \, e^{ib(\zeta\!-\!\xi)}(w^x\!\!+\! iw^y) (\!\zeta\!),
 \:\: \hbup\!=\!\hbxp{}'\!,\\[12pt]
\displaystyle  \hat u^z\!=\!\hat z'\!=\!\frac {\hbup{}^2}{2}\!=\!\E\!=\!\hat\gamma\!-\!1, \quad
\hat z(\xi)\!=\!\!\int^\xi_0\!\!\!\!  d\zeta \:\frac {\hbup{}^2\!(\zeta)}{2}.
\ea   \label{SolEqBz}
\eea
(\ref{SolEqBz}) reduces to the solution of \cite{KolLeb63,Dav63} 
for monochromatic $\Bep$. This leads to {\it cyclotron autoresonance} if 
$-b\!=\!k\!=\!\frac{2\pi}{\lambda}\!\gg\!\frac 1l$: for circular polarization \ $w^x(\xi)\!+\!iw^y(\xi)\!\simeq\!  e^{ik\xi} {\rm w}(\xi)$,   
\bea
( \hat x \!+\! i \hat y)(\xi)\simeq  i  W\!(\xi)e^{ik\xi} ,  \quad 
W(\xi)\!\equiv\!\!\int^\xi_0\!\!\!\!  d\zeta \, {\rm w}(\zeta)\!>\!0     \nonumber                      \label{ApprSolEqBz-2}
\eea
where ${\rm w}(\xi)\! \equiv\! q\epsilon(\xi)/kmc^2$; clearly $W(\xi)$ grows with $\xi$. 
In particular if $\Bep(\xi)\!=\!\0$ for $\xi\!\ge\! l\!\equiv$, then for such $\xi$
$$
\hat z'(\xi)\!\simeq\!  \frac {k^2}2 W^2(l)\!\simeq\!2\E_f, \quad
\frac{|\hbxp{}'(\xi)|}{\hat z'(\xi)}\!\simeq\!\frac 2 {k W\!(l)}\!\ll\! 1; 
$$

\subsection{Case \ $\bE_s=E_s^z\bk$, \ $\bB_s=\0$}
\label{Ezconst}

Then the solution (\ref{SolEqBzEz}) reduces to \ $\hat s(\xi)=1\!-\! \kappa\,\xi $,
\be
 ( \hat x +i \hat y)(\xi)=\!\!\int\limits^\xi_0\!\!\!  dy \,
 \frac{(w^x\!\!+\!iw^y) (y)} {1\!-\!\kappa y},\quad
\hat z(\xi)=\!\!
\int\limits^\xi_{0}\!\!\!\frac {dy}2 \left\{\!\frac {1\!+\! \hat v(y)}{[1\!-\!\kappa y]^2}\!-\!1\right\}\!;
\label{Ezcost}
\ee
If $\Bep$ is slowly modulated
the energy gain (\ref{EnergyGain}) $\E_f$ is negative if  $\kappa\!>\!0$, positive if   $\kappa\!\le\!0$
and has a unique maximum point $\kappa_{\scriptscriptstyle M}<0$ if $\epsilon(\xi)$ has a finite support with
a unique maximum. Here is an acceleration device based on this solution: 
at $t=0$ the particle initially lies at rest with $z_0\!\lesssim\! 0$, just at the left of a metallic grating $G$ contained in the
 $z\!=\!0$ plane and set at zero electric  potential;
another metallic plate $P$  contained in a  plane $z\!=\!z_p\!>\!0$  is set at electric potential $V=V_p$.
A short laser pulse $\Bep$ hitting the particle boosts it into the latter region
through the ponderomotive force; choosing $qV_p\!>\!0$ implies $\kappa\!=\!-qV_p/z_p mc^2\!<\!0$, and a 
backward longitudinal electric force $qE^z_s$. If  $qV_p$ is large enough, then $z(t)$ will reach a maximum
smaller than $z_p$, then is accelerated backwards and exits the grating
with energy $\E_f$ and negligible transverse momentum. 
A large $\E_f$ requires extremely large $|V_p|$, far beyond the material breakdown threshold,
what prevents its realization as a static field (namely, sparks between $G,P$ would arise and rapidly reduce $|V_p|$).
A way out is to make the pulse itself generate such large  $|E_s^z|$ within a plasma at the right time so as to induce
the {it slingshot effect}, as sketchily explained at the end of next section.

\section{Plane plasma problems}

Assume that the plasma is initially in hydrodynamic conditions with all initial  data [velocities, densities $n_h$, EM fields of the form (\ref{EBfields})] not depending on $\bxp$. Then also the solutions for  $\bB,\bE,\bu_h,n_h,\Delta\bx_h\equiv \bx_h(t,\bX)\!-\!\bX$  (displacements)
do not depend on $\bxp$. \ Here $\bx_h( t,\!\bX)$ is the position at $t$
of the $h$-th  fluid  material element with initial position $\bX\!\equiv\!(X,\!Y,\!Z)$; $\bX_h( t,\bx)$
is the inverse (at fixed $t$).
More specifically, we consider the impact of an EM plane wave with a pump of the type (\ref{aa'}.a) 
on a cold  plasma at equilibrium (figure below); the initial conditions are:
\begin{figure}[ht]
\includegraphics[width=8cm]{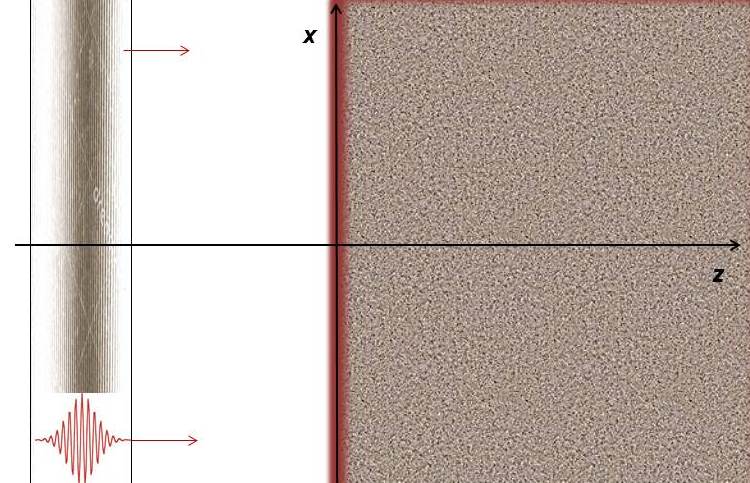}
\end{figure}
\bea 
\ba{l}
\bu_h(0,\bx)\!=\!\0,\qquad
n_h(0,\bx)\!=\!0\qquad \mbox{if }\: z\!\le\! 0, \\[8pt]
 j^0(0,\bx)\!=\!\sum\limits_h\!q_hn_h(0,\bx)  \!\equiv\! 0, \\[8pt]
\bE(0 ,\bx)\!=\!\Bep(-z),\quad \bB(0,\bx)\!=\!\bk\wedge\Bep(-z) +\bB_s.
\ea                                                       \label{incond0}
\eea
Then Maxwell eq.s $\nabla\!\!\cdot\!\bE\!=\!4\pi j^0$,
$ \partial_t E^z\!/c\!\!+\!4\pi j^z \!\!=\!(\!\nabla\!\!\wedge\!\bB)^z\!\!=\!0$ (the current density
is $\bj=\sum\limits_h\!q_hn_h\bb_h=\sum\limits_h\!q_hn_h\frac{\bu_h}{\gamma_h}$) imply  \cite{Fio14JPA}
\be
 \ba{l} E^{{\scriptscriptstyle z}}(t,\!z)=4\pi \sum_h q_h
\widetilde{N}_h[ Z_h(t ,z)], 
\ea            \label{expl}
\ee
where  $\widetilde{N}_h(Z)\!\equiv\!\int^{Z}_0\!\!d \zeta\,n_h(0,\!\zeta)$: \ 
we thus reduce by one the number of unknowns,  expressing $E^z$ in terms of the (still unknown) longitudinal motion.
$\bAp$ is coupled to the currents through $\Box\bAp=4\pi\bjp$ (in the Landau gauges). Including (\ref{incond0})
this amounts to the integral equation
\be
\bA\!^{{\scriptscriptstyle\perp}}\!\!-\!
\Ba\!^{{\scriptscriptstyle\perp}}\!-\!\frac {\bB_s}2\!\wedge\!\bx \!=\!
2\pi \!\! \int \!\!\!  d\!s   d\zeta\, \theta(ct\!\!-\!\!s\!-\!|z\!\!-\!\!\zeta|)
\theta(\!s\!) \,\bjp\!\left(\frac sc,\zeta\right).  
       \label{inteq1}
\ee
The right-hand side (rhs) is zero for $t\le 0$, because $t=0$ is the beginning  of the laser-plasma interaction. 
Within {\it short} time intervals $[0,t']$ (to be determined {\it a posteriori})  we can 
approximate \ $\bAp(t,z)\simeq \Bap(ct\!-\!z)\!+\!\frac {\bB_s}2\!\wedge\!\bx$; \ we also neglect the 
motion of  ions with respect to that of electrons.
Then the  Hamilton equations for the electron fluid with `time' $\xi$ and the initial conditions amount to (\ref{eqx})   and  
\be
\ba{l}
m c^2\hat s_e'(\xi,Z)=4\pi e^2 \left[
\widetilde{N}( \hat z_e ) \!-\! \widetilde{N}( Z)\right] + e(\hbxp_e{}'\!\wedge\!\hbBp_s)^z\!, \\[10pt]
m c^2 \hbup_e{}'(\xi,Z) =e\Bap{}' - e(\hat\bx'_e\!\wedge\!\hat\bB_s)^{\scriptscriptstyle \perp}\!,
\ea      \label{equps1}
\ee
\be
\hat\bx_e(0,\!\bX)\!=\!\bX, \quad \hat\bu_e(0,\!\bX)\!=\!\0\quad\Rightarrow \:\:\hat s_e(0,\!\bX)\!=\!1.     
\label{incond}
\ee
\smallskip
this is a family parametrized by $Z$ of {\it decoupled ODEs} which can be solved numerically.
The approximation on $\bAp\!(t,\!z)$ is acceptable as long as the so determined motion makes \ $|\mbox{rhs}(\ref{inteq1})| \!\ll\! |\Bap\!\!+\!\frac {\bB_s}2\!\!\wedge\!\bx|$; \ otherwise rhs(\ref{inteq1}) determines the first correction to $\bAp$; and so on.

\begin{figure}[ht]
\includegraphics[width=8.5cm]{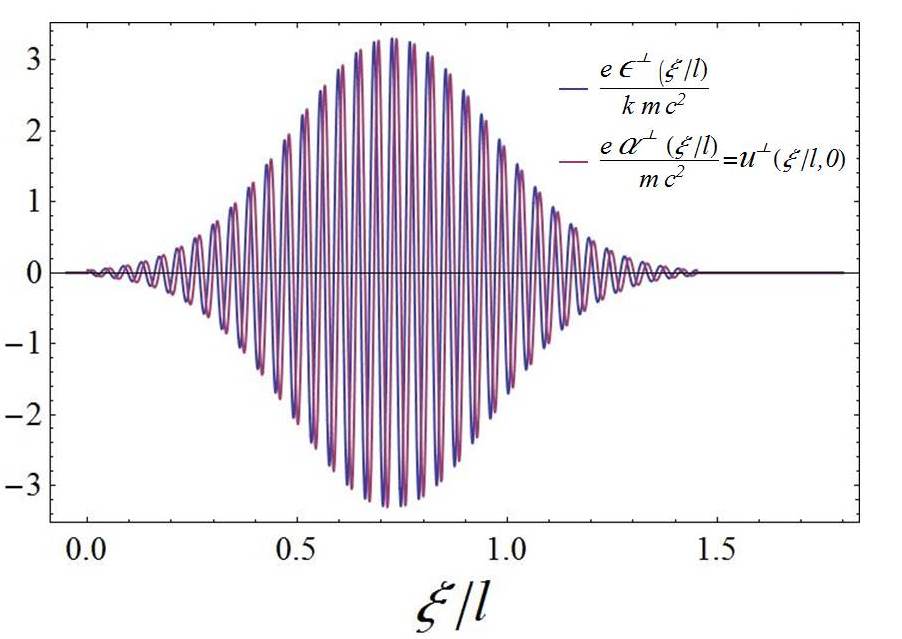}\\
\caption{Normalized  pump $\Bep$ with 
carrier wavelength $\lambda\!=\!0.8\mu$m, gaussian modulation $\epsilon(\xi)=a\exp[-\xi^2/2\sigma]$,
$ \sigma \!=\! 20\mu$m$^2$, \ $e a\lambda/mc^2\!=\!15$, average pulse intensity 
$10^{19}$W/cm$^2$, linear polarization.  $l\!\simeq\! 27\mu$m is the length of the interval  where the pump
amplitude $\epsilon$ overcomes the ionization threshold  for the cold gas (here helium)  yielding the plasma;
under such pulses the thresholds for $1^{st}$ and $2^{nd}$ ionization  are overcome (i.e. Keldysh parameters become
smaller than 1) almost simultaneously  \cite{Puk02,JovFedTanDeNGiz12}.}
\label{pulseR16gaussian}
\end{figure}

\begin{figure*}[ht]
\hskip0.5cm\includegraphics[width=15.5cm]{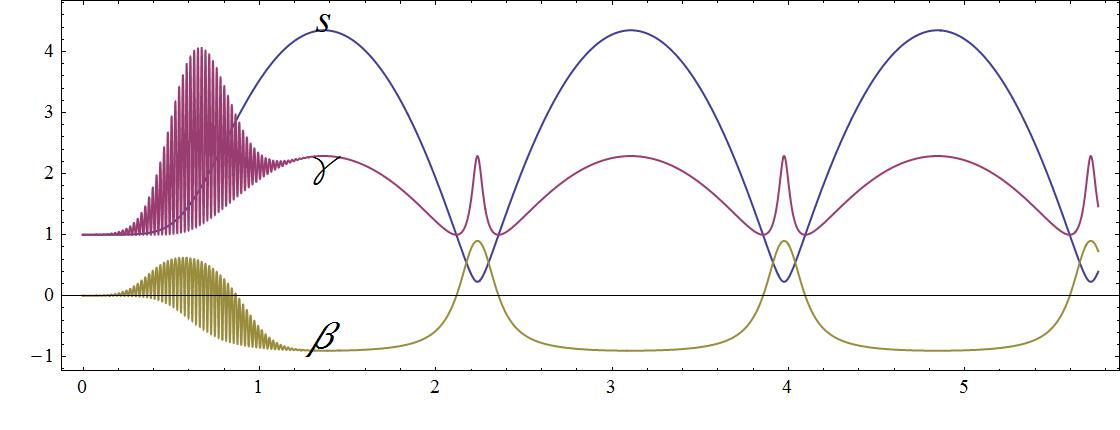} \\
\includegraphics[width=16cm]{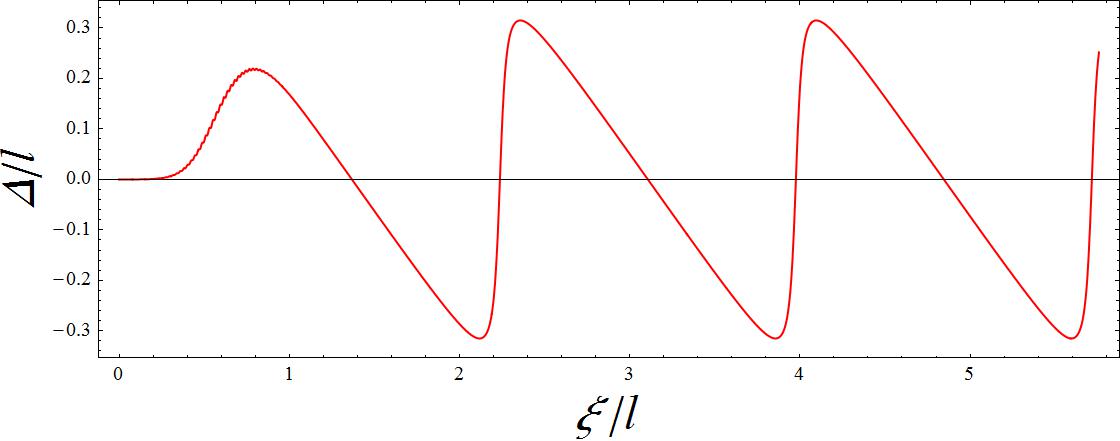} 
\caption{Solution of (\ref{heq1}-\ref{u_es_e}) corresponding to the pulse of fig. \ref{pulseR16gaussian},
 initial density $\widetilde{n_{e0}}(Z)\!=\!n_0\theta(Z)$,  $n_0\!=\! 2 \!\times\! 10^{18}$cm$^{-3}$.   
}
\label{graphs}
\end{figure*}

If $\bB_s\!=\!\0$, again (\ref{equps1})$_2$ is solved by $\hbup_e(\xi)\!=\! e\Bap(\xi)/mc^2$,  
while, setting $ v\!=\!\hat\bu^{{\scriptscriptstyle\perp}2}$, \ (\ref{eqx})$_2$, (\ref{equps1})$_1$ take   \cite{FioDeN16}
the form of (\ref{heq1r})
\bea
\Delta\hat z_e'\!=\displaystyle\frac {1\!+\! v}{2\hat s^2}\!-\!\frac 12, \quad
\hat s'_e\!=\frac{4\pi e^2}{mc^2} \left\{ 
\widetilde{N}[\hat z_e] \!-\! \widetilde{N}(Z) \right\}.  \label{heq1} 
\eea
If $n_e(0,\!\bX)\!=\!n_0\!=$const  for $Z\!\ge \!0$, 
then as long as $\hat z_e(\xi,Z)\!>\!0$ (\ref{heq1}), (\ref{incond}) reduce to 
the {\it same} Cauchy problem  {\it for all $Z$}:
\bea
&& \Delta '=\displaystyle\frac {1+v}{2s^2}\!-\!\frac 12,\qquad\quad
s'=M\Delta,\label{e1} \\ 
&&  \Delta (0)\!=\!0, \qquad\qquad\qquad   s(0)\!=\! 1\label{e2}
\eea
with $M \!\equiv\!\frac{4\pi e^2n_0}{mc^2}$. In fig. \ref{graphs} we show
the solution  if $\Bep$  is as in fig. \ref{pulseR16gaussian} and $n_0\!=\! 2\! \times\! 10^{18}$cm$^{-3}$;
$s(\xi)$ is indeed insensitive to the fast oscillations of $\Bep$ (see section \ref{A=A(t,z)}).
After the pulse is passed it becomes periodic: 
a plasma travelling-wave of spacial period $\xi_{\scriptstyle H}\simeq 49\mu$m follows the pulse.
The other unknowns are obtained through (\ref{hatsol}).
Replacing in rhs(\ref{inteq1}) we find that $\bAp\!\simeq\!\Bap$ is verified 
at least  for $t\!\le\! 5 \xi_{\scriptstyle H}/c$

\begin{figure*}
\includegraphics[width=7.9cm]{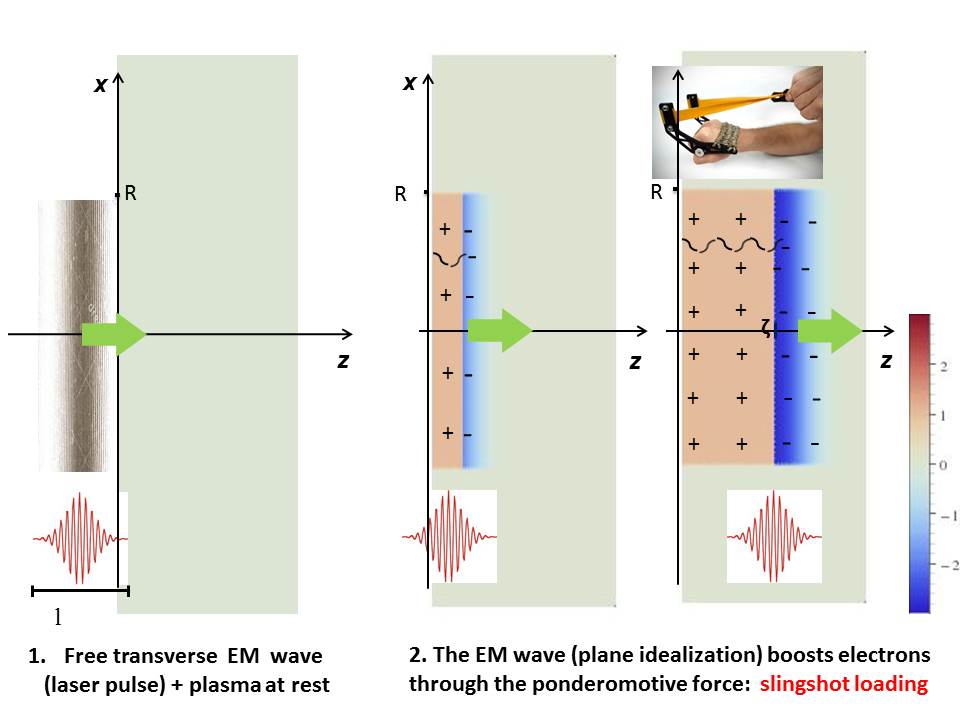}\hfill
\includegraphics[width=8.3cm]{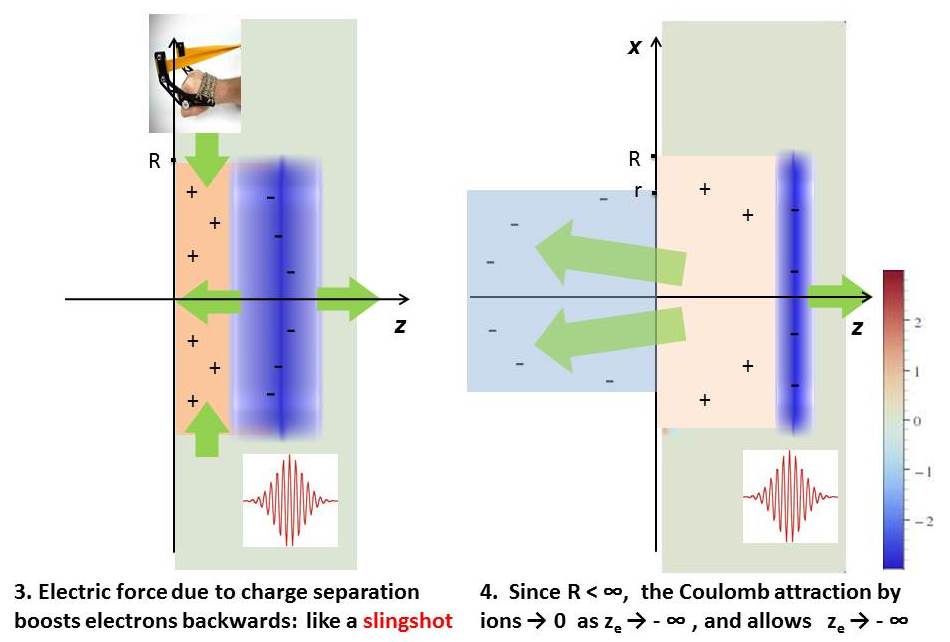}
\caption{Schematic stages of the slingshot effect.}
\label{StagesSlingshot}
\end{figure*}

The above  results  are based on a laser spot size $R=\infty$ (plane wave). When including corrections due to  
the finite $R$ (based on causality and heuristic estimates), they imply:
the impact of a very short and intense  laser pulse 
on the surface of a cold low-density  plasma (or gas,  ionized into a plasma by the pulse itself)
 may induce (for carefully tuned $R$), 
beside  a wakefield propagating behind the pulse \cite{Fio17c,Fio17b},
also  a \underline{backward acceleration and expulsion}   of surface electrons
\cite{FioDeN16,FioFedDeA14} ({\it slingshot effect}), as schematically depicted in fig. \ref{StagesSlingshot}.
For reviews see also \cite{FioActRic}.

\end{document}